\documentclass[doublecol]{epl2} 
\usepackage{amsmath}

\newcommand{\be}{\begin{equation}}
\newcommand{\ee}{\end{equation}}
\newcommand{\B}{h}

\title{Disordered Topological Insulators via $C^*$-Algebras}

\author{Terry~A.~Loring\inst{1} \and Matthew~B.~Hastings\inst{2}}
\shortauthor{Terry~A.~Loring and Matthew~B.~Hastings}

\institute{                    
  \inst{1} University of New Mexico - Department of Mathematics and Statistics,
  Albuquerque, NM 87131, USA\\
  \inst{2} Microsoft Research - Station Q, Elings Hall,
  University of California, Santa Barbara, CA 93106, USA
  and
  Department of Physics, Duke University, Durham, NC, 27708, USA\\
  }


\abstract{
The theory of almost commuting matrices can be used to quantify
topological obstructions to the
existence of localized Wannier functions with time-reversal
symmetry in systems with time-reversal symmetry and strong spin-orbit coupling.
We present a numerical procedure that calculates a $Z_2$ invariant using
these techniques, and apply it to a model of HgTe.
This numerical procedure allows us to access sizes
significantly larger than procedures based
on studying twisted boundary conditions.
Our numerical results indicate the existence of
a metallic phase in the presence of scattering between up and down spin
components, while there is a sharp transition
when the system decouples into two copies of the quantum Hall effect.
In addition to the $Z_2$ invariant calculation in the case
when up and down components are coupled, we also present a simple
method of evaluating the integer invariant in the quantum Hall case
where they are decoupled.
}

\begin{document}

\maketitle

The study of topological insulators is one of the most active areas of physics today.
Experimental and theoretical work has shown physical
realizations of time-reversal invariant insulators with strong
spin-orbit coupling in both two\cite{2d} and three dimensions\cite{3d} and
a complete classification of different insulating phases has
been recently obtained using methods of Anderson localization\cite{ludwig} and,
more generally, $K$-theoretic techniques\cite{kitaev}.

However, numerically it is difficult to determine the $Z_2$ invariants that
are signatures of topological insulating phases.
For systems with translational invariance, one can
study the bundle over the momentum torus\cite{momentum}, while for systems
without translation invariance, Essin and Moore\cite{essinmoore}
were able to study the phase diagram of a graphene model by studying the
model over a flux torus corresponding to twisted
boundary conditions.  Unfortunately, the flux torus approach is very
computationally intensive: for each disorder realization, the
Hamiltonian must be diagonalized once for each point on a discrete grid
on the flux torus, and then the connection on the
torus must be computed.  This limited the study to small 
systems, with at most 64 sites.

In this paper, we present a different approach to calculating 
a $Z_2$ invariant, based on ideas in $C^*$-algebras, in particular the
$K$-theory of almost commuting matrices.  We present a fast 
numerical algorithm based on these ideas.  Computing the invariant
requires a {\it single} diagonalization of the Hamiltonian, 
matrix function calculations on matrices at most half the
size of the Hamiltonian, and finally the calculation of
the Pfaffian of a real anti-symmetric
matrix that is at most the size of the Hamiltonian.
The most costly step is a single
diagonalization, allowing us to study significantly larger 
samples, up to 1600 sites.

Our invariant serves the same fundamental purpose as the invariant
used in \cite{essinmoore}---proving that for certain low-energy bands
it is impossible to find well-localized Wannier functions with
time-reversal symmetry.  These
invariants are most likely equivalent, but that is another topic
\cite{NumericalPractice}.

We apply this method to a model of HgTe \cite{2d}, including the
additional term $H_\mathrm{BIA}$ in \cite{hgtemodel} which breaks 
inversion symmetry.
The Hamiltonian we use is $\mathcal H$ of
\cite{hgtemodel}, with an additional on-site disorder term.  This can be written as
\be
{\mathcal H}=\begin{pmatrix} h & 0 \\ 0 & h^* \end{pmatrix} + H_\mathrm{BIA} + V,
\ee
where in the notation of \cite{hgtemodel}
\be
H_\mathrm{BIA} = \left(\begin{array}{cccc}
 &  &  & -\Delta\\
 &  & \Delta\\
 & \Delta\\
-\Delta\end{array}\right)
\ee
is a $k$-independent term that couples up and down spin components,
and $V$ is a random disorder term,
diagonal in spin and band indices, described below.
We map out part of the phase diagram, focusing on the
transition with and without coupling between up and down spins.

\section{Band Projected Position Operators and  
Wannier Functions in Systems Without Time-Reversal Symmetry}

We describe first an integer invariant in the case of systems 
without time-reversal symmetry.  We
largely follow~\cite{almost}, which described an integer 
invariant for two-dimensional systems on the sphere
or torus, and a $Z_2$ invariant in the case of a sphere.  In the next section,
we present a $Z_2$ invariant for systems on a torus.

Consider a lattice Hamiltonian $H$ for a single particle, 
tight-binding model on the surface of a torus.  We assume that the hopping is
short-range, so that the matrix element $H_{ij}$ is small 
if sites $i$ and $j$ are far separated on the torus.  Let
$P$ be the projector onto the states below the Fermi 
energy $E_F$.
If the Fermi energy lies in a spectral gap, then one
can show using the locality of $H$ that $P$ is also 
local: $P_{ij}$ is small if $i$ and $j$ are far separated.  In fact,
$P$ is still local even if the Fermi energy only lies in a {\it mobility} gap.
We parametrize the position of
a given site $i$ on the surface of the torus by two angles, $\theta_i$
and $\phi_i$, between $0$ and $2\pi$.  Introduce two Hermitian
matrices, $\Theta$ and $\Phi$.  These are both diagonal 
matrices, with matrix elements $\Theta_{ii}=\theta_i$ 
and $\Phi_{ii}=\phi_i$.

Using a basis of eigenstates of $H$, we can conjugate
the {\it band-projected position matrices} $P \exp(i \Theta) P$ 
and $P \exp(i \Phi) P$ by a single unitary to produce block matrices
\be
P \exp(i \Theta) P \sim \begin{pmatrix} 0 & 0 \\ 0 & U \end{pmatrix}, \;
P \exp(i \Phi) P \sim \begin{pmatrix} 0 & 0 \\ 0 & V \end{pmatrix}, \;
\ee
where the first block corresponds to the empty states 
(those projected onto by $1-P$) and the second block corresponds to the
filled states (those projected onto by $P$).

The matrices $U$ and $V$ 
are approximately unitary ($ U^\dagger U \approx I $
and $ V^\dagger V \approx I $) 
whenever $E_F$ lies in a mobility gap and the system is large.  To 
see this, note that if $P$ is local, then $P$ almost commutes with
$\exp(i \Theta).$ More precise results for a 
system on a sphere are in \cite{almost}.
Further, $U$ and $V$ almost commute if $E_F$ is in a mobility gap.

Now we ask: given two matrices, $U$ and $V$, 
which almost commute and which are almost unitary, does there exist
a pair of matrices $U',V'$ which exactly commute, are exactly 
unitary, and which are close to $U$ and $V$?  This question of
approximating almost commuting matrices by exactly 
commuting matrices is an old problem in $C^*$-algebras\cite{Halmos}.

Consider the quantity
\be
\label{Zinvar}
{\rm tr}(\log(V U V^\dagger U^\dagger)) \equiv
2 \pi i m + r,
\ee
where $m$ and $r$ are real.
We have $\exp(2\pi i m+r)$ equal to
${\rm det}(V U V^\dagger U^\dagger)$.  This determinant
is real and positive (since
${\rm det}(U)=\overline{{\rm det}(U^\dagger)}$)
so $m$ is an integer.  
This integer $m$ is precisely the topological invariant called
the {\it Bott index.}  The 
approximation by exactly commuting matrices is possible
if and only if $m=0$\cite{ELP98}.

If $U$ and $V$ almost commute and are close
to unitary, $V U V^\dagger U^\dagger$ is close to the identity.
All of its eigenvalues stay away from the
branch cut of the logarithm (chosen on the negative 
real axis), which is why the integer $m$ is a topological 
invariant.  If $U$ and $V$ exactly commute then $m=0$.

The physical importance of the Bott index is that when it
is non-zero, it is not possible
to find a complete, orthonormal basis of  localized
functions (so-called Wannier functions) spanning the occupied
states\cite{almost}.

This invariant  can be  computed 
very quickly.  We perform a full diagonalization of the
Hamiltonian to form the projector $P$ onto states below 
$E_F$.  We then construct $U$ and $V$ and calculate the
trace of the logarithm
in eq.~(\ref{Zinvar}) from the eigenvalues
of $VUV^\dagger U^\dagger$.
The matrices $U$ and $V$ are of dimension equal to
the number of occupied states.  This is at most half the
size of the dimension of $H.$

\begin{figure}
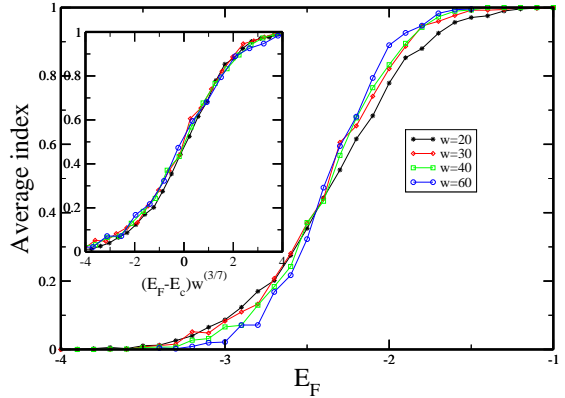

\onefigure[scale=0.30]{one.eps}
\caption{Average index as a function of the Fermi level $E_F$, for $w=20$ 
(1200 samples), $w=30$ (520 samples), $w=40$ (800 samples), 
$w=60$ (470 samples).  Inset: scaling collapse with $\nu=7/3$\cite{nu}.
\label{figZ}}
\end{figure}

Fig.~\ref{figZ}, shows the results of this invariant on a model 
of HgTe studied in \cite{hgtemodel}, with the up and
down spin decoupled ($\Delta=0$).  We chose constants
$A=1,B=-1,D=0,M=-2$,
which sets the irrelevant terms of order $k^2$ to zero.
The system had linear size $w$ giving $w^2$ sites and
$2w^2$ states per spin
component, and we added a diagonal disorder term on each 
site chosen uniformly from the interval $[-4,4]$.
The transition sharpens as $w$ increases.  In the inset,
we plot a scaling collapse with the exponent $\nu=7/3$ from
\cite{nu}.  This collapse suggests that the transition
is sharp in the thermodynamic limit.

\begin{figure}
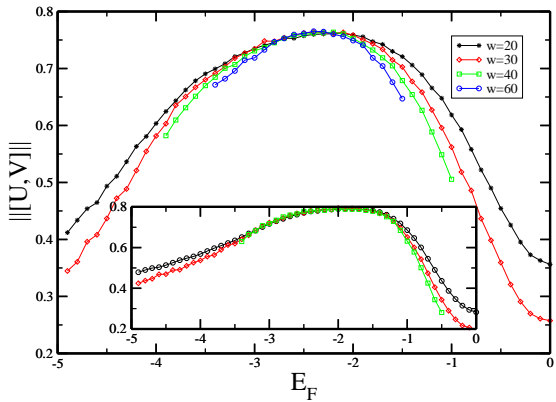

\onefigure[scale=0.30]{two.eps}
\caption{Norm of commutator $[U,V]$ as a function of $E_F$.  
Inset: same plot, for $\Delta=0.64,M=-1$, with $300$ samples 
for $w=40$, and $>400$ samples for smaller $w$.
\label{figUV}}
\end{figure}

One can also use the commutator of $U$ and $V$ as a signature 
of localization.  In fig.~\ref{figUV}, we plot this commutator as a function
of  $E_F$ for various system sizes.  One may see that as $w$ 
increases, the commutator gets smaller away from the transition.
In the inset, we show the same plot for a system with $\Delta=0.64,M=-1$,
to show the existence of
a metallic phase in this case.  

Finally, we mention the integer invariant on a sphere.
To describe a sphere, we introduce three coordinate matrices 
$X_1,X_2,X_3$, labeling the positions of the sites in 
three-dimensional space, with the radius of the sphere
set to one so that
$\sum_a X_a^2=I$.  We then project these matrices into the 
band of occupied states to form three matrices which we
call $H_1,H_2,H_3$.  Then if $E_F$ is in a mobility gap,
$\sum_i H_i^2$ is close to the identity and $H_i$ almost commutes with $H_j$.
We form the matrix
\be
B(H_1,H_2,H_3)=\sum_i H_i \otimes \sigma_i,
\ee
where $\sigma_i$ are the Pauli spin matrices.  One may show 
that $B^2$ is close to the identity if the $H_i$ almost commute
and $\sum_i H_i^2$ is close to the identity.  The integer invariant, 
also called the Bott index,
is half the difference between the number of positive and 
negative eigenvalues of $B$.

The reader may also ask about invariants on a system with 
open boundary conditions, such as a disk\cite{topofermi},
where there are two almost commuting band projected position 
matrices, $PXP,PYP$.  In this
case, there is no topological obstruction to approximating 
almost commuting matrices
by exactly commuting matrices\cite{lin,macmc}; physically, 
we understand this as if the system on a disk is
in a topologically nontrivial phase then there will be 
gapless boundary modes and hence $PXP,PYP$ will not almost commute.

\section{An Obstruction to Localized Wannier Functions with Time-Reversal
Symmetry}

We now consider invariants in systems with time reversal symmetry.
We begin with the invariant for matrices that almost represent a
sphere constructed in \cite{almost}.  This theory applies to all
triples $(H_1, H_2, H_3)$ of self-dual Hermitian matrices for which
\be
\| [ H_r, H_s ] \| \leq \delta ,\quad \| I - \sum H_r  \| \leq \delta
\label{sphere-rep}
\ee
with a small $\delta.$  We need not convert a lattice over the
torus to a lattice over the sphere; we compute the band-compressed
periodic position operators and apply matrix functions to
create matrices that satisfy eq.~(\ref{sphere-rep}).

Consider the matrix $\widetilde{B}(H_1,H_2,H_3),$ which we define as
\be
\tfrac{i}{2}(1+i \sigma_y \otimes \sigma_y) B(H_1,H_2,H_3) (1-i\sigma_y \times \sigma_y),
\ee
where $\sigma_y\otimes \sigma_y$
is a product of the $\sigma_y$ matrices on the physical spin of the particle
and on the pseudospin introduced to define $B$.
The resulting matrix is real and anti-symmetric. The sign of its Pfaffian
represents a potential $Z_2$ obstruction in the group $Z_2=\{-1,1\}.$

We call this index the {\it Pfaffian-Bott index.} This index being $-1$
is first of all an obstruction to our being able to
approximate the $H_r$ simultaneously by exactly commuting self-dual matrices.
The analysis in \S V.B of \cite{almost} is valid in the
self-dual case, and it shows that this is also an obstruction to
finding exponentially localized Wannier functions with time-reversal
symmetry.

The spectrum of $\widetilde{B}(H_1,H_2,H_3)$ is pure imaginary and has
a gap at zero.  The smaller the commutators between the $H_r$ the larger the
gap, and the larger the distance must be to exactly commuting
self-dual matrices.

We prefer to study a physical system on a torus in order to have a
regular lattice without defects.
The $Z_2$ invariant for the torus is not as simple as eq.~(\ref{Zinvar}).
We compute the matrices
$U$ and $V$ as above; these matrices $U,V$ will be self-dual
if we are careful how we diagonalize the Hamiltonian.  We need
to diagonalize via a symplectic matrix, found using a variation
of an old algorithm~\cite{algo} designed for
Hamiltonians with time-reversal symmetry.

We select a degree-one, continuous mapping from the torus to the sphere, a map 
$(\theta,\phi) \mapsto(x_1,x_2,x_3) $
with
\begin{align}
x_1 &= f(\phi) \notag\\
x_2 &= g(\phi) + h(\phi)\cos{2\pi i \theta} \\
x_3 &= h(\phi)\sin{2\pi i \theta} \notag
\end{align}
We then
apply this map to $U$ and $V$ by defining
\begin{align}
H_1 &= f(V) \notag \\
H_2 &= g(V) + \tfrac{1}{4} \{h(V), U^\dag\} + \tfrac{1}{4} \{ h(V), U \}
\label{Hr-naive} \\
H_3 &=  \tfrac{i}{4} \{h(V),U^\dag\} - \tfrac{i}{4} \{h(V), U \} \notag
\end{align}
and then compute the $Z_2$ invariant of $H_1,H_2,H_3$.  The anticommutators
ensure that the $H_r$ are self-dual.

Expressing the torus invariant in terms of
the sphere invariant requires computing with the matrix
$\widetilde{B}(H_1,H_2,H_3)$, which has dimension
twice as large as the number of occupied states,
and so is at most the size of the Hamiltonian.
We need only tridiagonalize the matrix
$\tilde B,$ which is real, further improving the speed.

There are issues with applying the formulas in eq.~(\ref{Hr-naive}).
The matrix functions of $V$ must be analytic functions, Laurent polynomials
or polynomials in $V$  and $V^\dag$ since we don't expect $V$ to
be normal.  That limitation is not compatible with defining a
degree-one map from the torus to the sphere.  This we overcome by
choosing $f,$ $g$  and $h$ that define a map to three-space that
is close to the sphere.

\begin{figure}
\onefigure[scale=0.36]{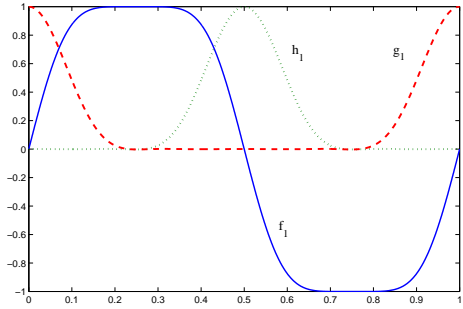}
\caption{Graphs of the $f_1,$ $g_1$ and $h_1$ as scalar functions
\label{threefunctions}}
\end{figure}

Our choice of $f$ is similar that that used in \cite{ExelLoring, almost}, only
with three derivatives vanishing at the extrema. See fig.~\ref{threefunctions}.
We define $f_1$, $g_1$ and $h_1,$ truncating the
Fourier series for $g_1$ and $h_1$ to produce
order-5 trigonometric polynomials:
\begin{align}
f_1(\phi) &= 
\tfrac{150}{128}\sin(2\pi \phi)+\tfrac{25}{128}\sin(6\pi \phi)+\tfrac{3}{128}\sin(10\pi \phi)
\notag \\
g_1(\phi) &\approx
\begin{cases}
0 & \tfrac{1}{4}\leq\theta\leq\tfrac{3}{4}\\
\sqrt{1-\left(f(\phi)\right)^{2}} 
& -\tfrac{1}{4}\leq\theta\leq\tfrac{1}{4}\end{cases}
\label{trigMethod} \\
h_1(\phi) &\approx
\begin{cases}
\sqrt{1-\left(f(\phi)\right)^{2}} 
& \tfrac{1}{4}\leq\theta\leq\tfrac{3}{4}\\
0 & -\tfrac{1}{4}\leq\theta\leq\tfrac{1}{4}\end{cases} \notag
\end{align}

\begin{figure}
\begin{center}
\onefigure[scale=0.32, trim = 45mm 00mm 10mm 20mm]{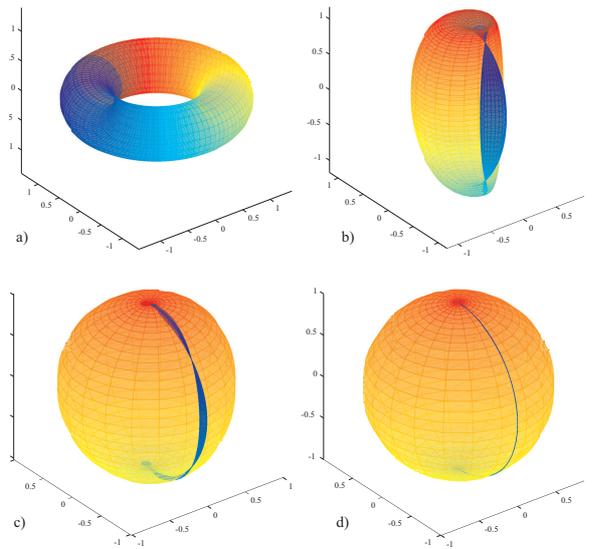}
\caption{Panel d) indicates using $f_1,$ $g_1$ and $h_1$ to map the torus
(panel a)) close to the sphere. Panels b) and c) show the result of
truncating the three functions to
degrees one and three. \label{spherePic}}
\end{center}
\end{figure}

Our invariant is most meaningful when $\|V^\dag - V^{-1}\|$ is small,
so we replace $g_1(V)$ by a star-polynomial $g(V)$ by
$g_1(V) = \sum _{n=-5}^{5} b_n V^n \approx g(V) = \sum _{n=-5}^{5} b_n V^{(n)}$
where $V^{(n)}$ equals $V^n$ for positive $n$ and $(V^\dag)^{-n}$ for
negative $n.$  This is much faster.  Applied to scalars
of unit modulus, there is no difference between $g_1(z)$  and $g(z).$
We do a similar replacement of inverse by adjoint to define $f(V)$
and $h(V).$
The underlying map from the two sphere is illustrated by 
fig.~\ref{spherePic}.

Based on earlier experience~\cite{ExelLoring}, we used also
an alternative method.  We computed the polar decomposition $V=V_0|V|$
of $V$ and a symplectic diagonalization $ W D W^\dag$ of $V_0.$  This
allowed discontinuous functions
\begin{align}
f(\phi) &= 1-2\phi \notag\\
g(\phi) &= 0 
\label{logMethod}\\
h(\phi) &= \sqrt{1 - f(\phi)^2} \notag
\end{align}
as we can apply them to the diagonal of $D.$

Both methods work well, with the star-polynomial
method, eq.~(\ref{trigMethod}), being faster.  The log-based method,
eq.~(\ref{logMethod}), gave smaller values for the important numbers
$
\| [ H_r, H_s ] \|
$
and
$
\|   I - \sum H_r  \|.
$
The data in the plots were generated using the log-based method.
These matrix functions, and $O(n^3)$ algorithms for Pfaffians
and symplectic diagonalization, will be discussed in \cite{NumericalPractice}.

The inability to approximate the $H_r$ by self-dual, exactly commuting
matrices implies the inability to approximate the $U$ and
$V$ by self-dual, exactly commuting matrices.
We have an obstruction to the existence of exponentially localized
Wannier functions with time-reversal symmetry in our original lattice
model on the torus.

\section{Results for a System With Time-Reversal Symmetry}

\begin{figure}
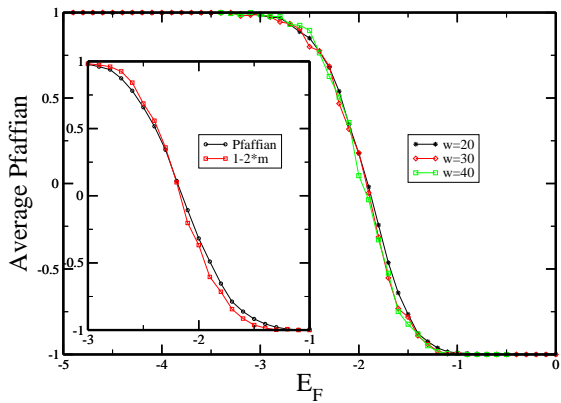

\onefigure[scale=0.30]{five.eps}
\caption{Average value of Pfaffian as a function of $E_F$,
for $w=20,30,40$ and $\Delta=0.64$.  Inset: comparison of
average value of Pfaffian for $\Delta=0,M=-1$ to $1-2*m$
for same system at $w=20$.
\label{figPf}}
\end{figure}

In fig.~\ref{figPf} we plot the average value of this Pfaffian as a function
of $E_F$ for a system using the model of HgTe studied in \cite{hgtemodel}.
We include spin-orbit coupling, which in the terms of \cite{hgtemodel}
means  $\Delta=0.64,M=-1$.  Unlike in fig.~\ref{figZ}, here the
plots do not continue to sharpen
as $w$ increases, suggesting the presence of a metallic phase.
When the $U$ and $V$ are far from commuting,
the spectral gap of
$\widetilde{B}(H_1,H_2,H_3)$ is small and the value
of the Pfaffian-Bott index is less
meaningful.  The wide plateau in fig.~\ref{figUV} inset
is perhaps a stronger indication of a metallic phase. 
The fluctuating index implies that
the phase is metallic.  
We are not certain of the interpretation of
the index in the metallic regime, but we note
that the index is a well-defined topological invariant
so long as the Green's function has a sufficiently fast power-law decay.

In fig.~\ref{figPf} inset, we compare for the case $\Delta=0$
the average value of the $Z_2$ invariant to the $Z$
invariant for the same system, showing
that the average value of the Pfaffian-Bott index is close to $1-2*m$.
Recall that nonzero Bott index and Pfaffian-Bott index not $1$
both indicate topological insulators.

\begin{figure}
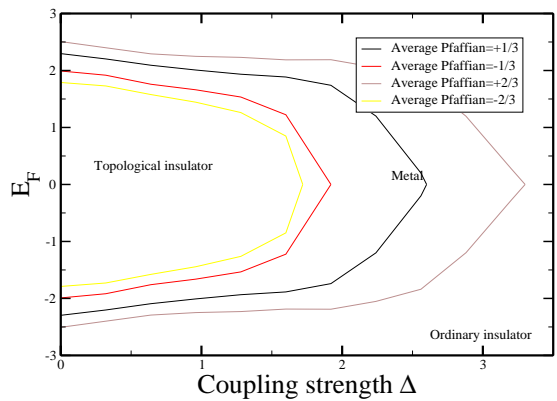

\onefigure[scale=0.30]{six.eps}
\caption{For given $\Delta$, plot of which $E_F$ gives average index
equal to $\pm 1/3,\pm 2/3$ for $w=20$.
\label{figphasediagram}}
\end{figure}

In fig.~\ref{figphasediagram}, we show the phase diagram as a function of $\Delta$ with
$M=-1$; in fact, $\Delta$ cannot be varied in experiment,
but the ratio $\Delta/M$ can be as $M$ depends on sample
geometry\cite{hgtemodel}.  For $w=20$, we computed which values
of $E_F$ gave average index equal to $\pm 1/3$ or $\pm 2/3$.
Due to the finite $w$, even at $\Delta=0$ there is a non-zero
distance between the curves.
In the thermodynamic limit, the insulator
has average Pfaffian $\pm 1$, and
the metal has absolute value of average Pfaffian less than $1$.
Finite size simulations cannot identify the insulating
regime with certainty, but these simulations suggest that
the topological insulator, where the index is always $-1$, is
located a short distance inside the $-2/3$ curve and
the ordinary insulator is located a short distance outside
the $+2/3$ curve.
The curves with $E_F>0$ were obtained by mirroring the
curves with $E_F<0$; while our index is not obviously
particle-hole symmetric, numeric tests indicate that
the average Pfaffian is in fact close to invariant
under this reflection.

Interestingly, for $\Delta \approx 2$, the system has a delocalized
phase near zero energy, but there is no topological insulating phase
at this value of $\Delta$ and the phase instead separates two ordinary
insulating phases.  Rather, the delocalized phase results from the
proximity to the topological insulating phase at smaller values of
$\Delta$.

\section{Relation of Bott Index to Hall Conductance}
Finally, we relate the Bott index, in the case without
time reversal symmetry, to the Hall conductance.  We do this by
relating the Bott index to an invariant\cite{ak} which
describes the response of charge
to a magnetic field.
Suppose $H$ has a spectral gap.
Then, we can ``spectrally flatten'' $H$, to define a new
matrix $H'$ whose eigenvalues are all equal to $\pm 1$, which is
still local in the sense that the matrix elements of $H'$ are
exponentially small between far separated sites, and such that
the space spanned by negative energy states of $H'$ is the
same as that spanned by occupied states of $H$.
From now on in this section, when
we write $H$ we in fact mean the spectrally flattened
Hamiltonian $H'$.  

We add a pseudospin degree of freedom with a Zeemann coupling
to a magnetic monopole in the center of the sphere,
giving the Hamiltonian
\be
\label{hcoupled}
\tilde H=H\otimes I + \B \sum_a X_a \otimes \sigma_a,
\ee
with the radius of the sphere equal to one so $\sum_a X_a^2=I$.  Then,
if the system is large, the
lattice spacing $a$ is small, and we can choose $\B$ such that
$a \ll \B \ll 1$.
Let $H$ have $M$ negative eigenvalues.  Then
$\tilde H$ has $2M$ negative eigenvalues since $\B\ll 1$.  
Since $a\ll \B$, if the pseudospin
is aligned with the magnetic field, and hops to a nearby site within distance
$a$, the angle between the pseudospin and the magnetic field on the new
site is small.  So for $a\ll \B$, we can do describe
the dynamics semi-classically: the particle hops on the lattice,
with its pseudospin following the magnetic
field.  In this semi-classical limit, the negative eigenvalues are all close
to $-1\pm \B$.

The invariant we will consider is the difference between the number of
negative eigenvalues close to $-1+\B$, which we denote
$M_+$, and the number of negative eigenvalues close to $-1-\B$, which
we denote $M_-$, where $M_++M_-=2M$.  We compute $M_+-M_-$ in two
different ways.

First, we compute it semi-classically.  Consider the eigenvalues close
to $-1+\B$.  If the pseudospin is aligned with the field, the Berry phase
for transport of the pseudospin leads to an effective orbital magnetic field,
as if there were a magnetic monopole inside the sphere.
We will write $P=(1-H)/2$ for the projector onto the negative energy states of $H$.
We write $P(+)$ for the projector onto the negative energy state of a modified
Hamiltonian, $H(+)$, where $H(+)$ is the Hamiltonian $H$ modified by an
orbital coupling to a magnetic monopole.  We write $P(-)$ for the projector
onto the negative energy states of a Hamiltonian $H(-)$, which has the opposite
sign monopole.  So,
$M_+-M_-={\rm tr}(P(+)-P(-))$.  This quantity
${\rm tr}(P(+)-P(-))$ has been suggested by \cite{ak}.

However, we can also compute $M_+-M_-$ using degenerate perturbation
theory.  Since $\B\ll 1$, we can ignore matrix elements of
$\B \sum_a X_a \otimes \sigma_a$ which couple negative and positive energy
states of $H\otimes I$ and
and just consider the operator $\B \sum_a X_a \otimes \sigma_a$ projected onto
the negative energy space of $H \otimes I$.  This operator equals
$\B \sum_a P X_a P \otimes \sigma_a=\B\cdot B(PX_1,PX_2,PX_3P)$.
Thus, in this semi-classical limit the Bott index equals the  difference
${\rm tr}(P(+)-P(-))/2$.

While the invariant of \cite{ak} may be unfamiliar, in fact physically it
simply describes the response of charge to flux.  In a quantum Hall
system, the adiabatic addition of flux increases the charge by an
amount proportional to the Hall conductance\cite{laughlin}.  
It is possible in this
way to equate the Bott index we consider with the Hall conductance of
a system on a sphere in the limit of large system size.

We now turn to the torus.  
The relation between Hall conductance and the Bott invariant
as in eq.~(\ref{Zinvar}) will be investigated in \cite{NumericalPractice}. 
The calculation in this
section suggests that we
want a map from the torus to the sphere such that the
pullback of the connection on $CP^1$ onto $T^2$ gives a connection 
with constant curvature
to give a constant effective orbital magnetic field on the torus 
with a single monopole. 
No such map exists but
the map we used in the log-method, eq.~(\ref{logMethod}),
gives constant curvature except
at the singularity $\phi=0$.

%SOME DISCUSSION OF CLASSIFICATION OF PAIRS OF UNITARIES, HIGHER DIMENSIONS, ETC???

\section{Discussion}
We have presented an approach to calculating $Z_2$ invariants in 
time-reversal invariant
systems, allowing us to study larger system sizes than previously.

In a numerical study of a model for HgTe, we observed
a metallic phase around the transition when the spin
components are not decoupled, as
expected from the existence of metallic phases in disordered
systems with spin-orbit scattering due to anti-localization\cite{symplectic}.

We use methods of non-commutative topology, not 
noncommutative geometry\cite{NCgeomQHE}, to define our indices.
While no precise boundary exists between these fields
in the mathematics literature, our general approach is to
discard most of the information in the original Hamiltonian, 
only retaining the band-projected position operators.  This keeps
topological properties while discarding
``metric information'' \cite{NCgeomSurvey} used
in the noncommutative geometry approach.  By retaining the
minimum of information, we obtain an efficient, practical
computational procedure for these indices.  
The topological invariance of these indices follows from
standard linear algebra, although the tighter
estimates in \cite{almost} require recent
results in matrix theory.

Two distinct approaches to investigating topological phases
are common.  One is via index and the other
is via transport.  In this paper we have taken the approach of
studying an index.  
In contrast to flux-torus approaches to $Z_2$ index, such as Essin-Moore,
our approach allows the numerical study of much larger systems.
For systems with coupling between spin components,
studying the Hall conductance as in \cite{transportGrothHgTe} does
not work, so the transport method requires boundaries.
In contrast, on such systems our method works without
the introduction of boundaries.
Although the transport approach allows larger system size,
introducing boundaries may increase finite size effects. Further,
the conductance in a transport measurement is not
exactly quantized in finite size systems, while the index approach
returns an integer topological invariant even for finite sizes.  Another
interesting ability of the index approach is in studying effects such
as we have seen near $\Delta=2$.  There, the delocalized phase has a
slightly negative average index (roughly $-0.3$) at $E_F=0$, implying
that the index detects the nearby presence of a topological insulator.
Ideally, we want
access to all properties of a topological insulator: bulk conductance,
conductance of boundary modes, and index.  We expect that a
combination of transport and index techniques will be needed to obtain
a full understanding of these systems in the presence of strong
disorder.

Since our approach does not use twisted boundary conditions, we
maintain the time reversal symmetry of the problem throughout the 
calculation.  This leads to a philosophical advantage of our
approach: rather than studying the response of the system to a 
perturbation that breaks the symmetry (the twisted boundary conditions),
we directly study an intrinsic property of the given Hamiltonian, 
in  particular allowing us to show obstructions
to localized Wannier
functions for that Hamiltonian.

\acknowledgments
MBH thanks M. Freedman, A. Kitaev, C. Nayak and X.-L. Qi for useful discussions.

\end{document}